\documentclass[a4paper,11pt]{article}
\usepackage{jheppub} 

\usepackage[T1]{fontenc} 
\usepackage{subcaption}
\usepackage{verbatim}
\usepackage{mathrsfs}
\usepackage{physics}
\usepackage{amsthm}
\usepackage{amsfonts}
\usepackage{mathtools}
\usepackage{xcolor}
\usepackage{enumerate}

\newtheoremstyle{breakthm}%
{}{}%
{\itshape}{}%
{\bfseries}{}
{\newline}{}

\theoremstyle{breakthm}
\newtheorem{theorem}{Theorem}

\renewcommand{\H}{\mathcal{H}}

\title{Reflected entropy is not a correlation measure}
\author[a]{Patrick Hayden}
\affiliation[a]{Stanford Institute for Theoretical Physics, Stanford University, Stanford, California 94305, USA}
\author[b]{Marius Lemm}
\affiliation[b]{Department of Mathematics, University of T\"{u}bingen, Geschwister-Scholl-Platz, 72074 T\"{u}bingen, Germany}
\author[c]{Jonathan Sorce}
\affiliation[c]{Center for Theoretical Physics, Massachusetts Institute of Technology, Cambridge, Massachusetts 02139, USA}

\abstract{
By explicit counterexample, we show that the ``reflected entropy'' defined by Dutta and Faulkner is not monotonically decreasing under partial trace, and so is not a measure of physical correlations.
In fact, our counterexamples show that none of the R\'{e}nyi reflected entropies $S_{R}^{(\alpha)}$ for $0 < \alpha < 2$ is a correlation measure; the usual reflected entropy is realized as the $\alpha=1$ member of this family.
The counterexamples are given by quantum states that correspond to classical probability distributions, so reflected entropy fails to measure correlations even at the classical level.
}

\makeatletter
\gdef\@fpheader{MIT preprint ID MIT-CTP/5532\vspace{3em}}
\makeatother
\notoc
\begin{document}
	\maketitle
\newpage

\section{Introduction}

In \cite{dutta2021canonical}, Dutta and Faulkner defined a function of bipartite quantum states called the reflected entropy.
For a density operator $\rho_{AB}$ describing a quantum state on the bipartite Hilbert space $\H_A \otimes \H_B,$ the reflected entropy is denoted $S_{R, \rho}(A:B).$
For finite-dimensional $\H_A$ and $\H_B,$ the reflected entropy can be described as follows.\footnote{While we will only be concerned with finite-dimensional systems in this paper, it is possible to define reflected entropy for certain infinite-dimensional density operators; this is discussed in \cite{dutta2021canonical}.}
One considers the operator $\sqrt{\rho_{AB}}$ as a member of the vector space $\H_A \otimes \H_B \otimes \H_A^* \otimes \H_B^*,$ where $\H^*$ denotes the dual space of $\H$.
Writing this in quantum mechanical notation as $\ket{\sqrt{\rho_{AB}}},$ it is easy to check that this is a pure state that purifies $\rho_{AB}$ in the sense that it satisfies
\begin{equation}
	\tr_{A^* B^*} \ketbra{\sqrt{\rho_{AB}}} = \rho_{AB}.
\end{equation}
The state $\ket{\sqrt{\rho_{AB}}}$ is called the canonical purification of $\rho_{AB}.$
The reflected density operator on the system $\H_A \otimes \H_A^*$, denoted $\phi^{(AB)},$ is defined as the reduced state of $\ket{\sqrt{\rho_{AB}}}$ on $\H_A \otimes \H_A^*,$ i.e.
\begin{equation}
	\phi^{(AB)}
		= \tr_{BB^*} \ketbra{\sqrt{\rho_{AB}}}.
\end{equation}
The reflected entropy is defined as the von Neumann entropy of this state,
\begin{equation}
	S_{R, \rho}(A:B)
		= S(\phi^{(AB)})
		= - \tr(\phi^{(AB)} \log \phi^{(AB)}).
\end{equation}
The reflected entropy belongs to a natural family of reflected R\'{e}nyi entropies $S_{R, \rho}^{(\alpha)}$ for $\alpha \in (0, 1) \cup (1, \infty),$ defined in terms of the standard R\'{e}nyi entropies $S_\alpha$ \cite{renyi1961measures} as
\begin{equation} \label{eq:renyi-reflected}
	S_{R, \rho}^{(\alpha)}(A:B)
		= S_\alpha(\phi^{(AB)})
		= \frac{1}{1-\alpha} \log\tr((\phi^{(AB)})^\alpha),
\end{equation}
with the ordinary reflected entropy being recovered in the limit $\alpha \to 1$.

In \cite{dutta2021canonical}, Dutta and Faulkner showed that the reflected entropy has a natural physical interpretation for certain states in holographic theories of quantum gravity.
In \cite{akers2020entanglement}, parametric gaps between the reflected entropy and the mutual information\footnote{Mutual information is defined below in equation \eqref{eq:MI}.} in certain settings were shown to be related to the presence of parametrically large amounts of tripartite entanglement.
The connection between reflected entropy and tripartite entanglement was studied further in \cite{zou2021universal}, where the authors considered the structure of states with reflected entropy equal to the mutual information, and in \cite{hayden2021markov}, where the authors gave an interpretation of the reflected entropy in terms of an information-theoretic reconstruction problem.
A great deal of work has been done to explore the physical properties of the reflected entropy in various systems of interest, including holographic models in \cite{jeong2019reflected, bao2019multipartite, akers2020entanglement, hayden2021markov, chu2020generalizations, bao2021entanglement, akers2022reflected, akers2022page, basu2022entanglement, akers2022reflected-2, dutta2022reflected} and many-body systems in \cite{kudler2020quantum, kusuki2020entanglement, kudler2020correlation, moosa2020time, bueno2020reflected-1, kudler2021entanglement, berthiere2021topological, kudler2021quasi, bueno2020reflected-2, camargo2021long, siva2022universal, liu2022multipartitioning, vardhan2023mixed, liu2023multipartite, zou2021universal}.
None of these explorations, however, answers a basic physical question: does the magnitude of $S_{R, \rho}(A:B)$ quantify the magnitude of the correlations between systems $A$ and $B$ in the state $\rho_{AB}$?
In the literature, it is commonly assumed that this is the case --- for example, in \cite{dutta2021canonical, hayden2021markov, akers2022reflected, basu2022entanglement, akers2022reflected-2, kusuki2020entanglement, kudler2020correlation, moosa2020time, berthiere2021topological, kudler2021quasi, liu2022multipartitioning, liu2023multipartite}, reflected entropy is referred to as a ``correlation measure,'' a ``measure of correlations,'' or a ``measure of entanglement.'' 
The purpose of this article is to show that the answer is no.

Let $\delta_{\rho}(A:B)$ be a generic rule for assigning nonnegative numbers to bipartitions of quantum states.
What does it mean for $\delta_{\rho}(A:B)$ to measure correlations between $A$ and $B$?
One basic requirement is that in a quantum state $\rho_{ABC}$ with marginal $\rho_{AB},$ there should be at least as much correlation between $A$ and $BC$ in the state $\rho_{ABC}$ as there is between $A$ and $B$ in the state $\rho_{AB}.$
So if $\delta$ is a measure of correlations, then it must satisfy the inequality
\begin{equation}
	\delta_{\rho}(A:BC) \geq \delta_{\rho}(A:B)
\end{equation}
for any tripartite density operator $\rho_{ABC}.$
A function satisfying this inequality is said to be \textit{monotonically decreasing under partial trace} or simply \textit{monotonic}.
One famous example of such a quantity is the mutual information,\footnote{For a proof that mutual information is monotonically decreasing under partial trace, see theorem 11.15 of \cite{nielsen2002quantum}.} defined by
\begin{equation} \label{eq:MI}
	I_{\rho_{AB}}(A:B) = S(\rho_{A}) + S(\rho_{B}) - S(\rho_{AB})
\end{equation}
where $S$ is the von Neumann entropy.

Up until now, it has been common in the literature to refer to the reflected entropy as a measure of correlations, despite it not being known whether it is monotonically decreasing under partial trace.
One primary reason is that in \cite{dutta2021canonical}, Dutta and Faulkner proved that the R\'{e}nyi reflected entropies [see equation \eqref{eq:renyi-reflected}] are monotonically decreasing under partial trace for $\alpha$ an integer greater than or equal to 2, and so they can reasonably be thought of as measures of physical correlations.
Furthermore, as observed in \cite{dutta2021canonical}, monotonicity of reflected entropy for holographic states follows as a simple consequence of the ``entanglement wedge nesting'' property of holography established in \cite{wall2014maximin}.
Monotonicity is not unique to holography --- in \cite{bueno2020reflected-2}, Bueno and Casini showed that reflected entropy satisfies monotonicity in certain free quantum field theories.
Finally, it is clear from the observations in \cite{akers2020entanglement, zou2021universal, hayden2021markov} that the reflected entropy has a meaningful relationship to entanglement, so it seems natural to believe that it should measure physical correlations.

This is not the case.
In the next section, we provide explicit counterexamples for monotonicity of the R\'{e}nyi reflected entropies in the range $0 < \alpha < 2,$ including the reflected entropy itself.
Our counterexamples are physical states on a system consisting of two qutrits and a qubit.
In fact, the states we construct are diagonal in a tensor product basis and hence can be described by classical probability distributions, so reflected entropy fails to measure correlations even in classical physics.\footnote{By a continuity argument, our result also implies that there exist non-classical states that violate monotonicity of reflected entropy. To see this, note that non-classical states can be found in an arbitrarily small neighborhood of any classical one and  the R\'{e}nyi reflected entropies are continuous under small perturbations of the state.}
For convenience, we state our main result clearly as the following theorem.

\begin{theorem}
	For any $\alpha \in (0, 2),$ there exists a density operator $\rho_{ABC}$ on $\H_A \otimes \H_B \otimes \H_C = \mathbb{C}^{3} \otimes \mathbb{C}^3 \otimes \mathbb{C}^2$ for which the $\alpha$-th R\'{e}nyi reflected entropy [see equation \eqref{eq:renyi-reflected}] satisfies
	\begin{equation}
		S_{R, \rho}^{(\alpha)}(A:BC) < S_{R, \rho}^{(\alpha)}(A:B).
	\end{equation}
\end{theorem}

We recall that Dutta and Faulkner proved in \cite{dutta2022reflected} that R\'{e}nyi reflected entropies at integer values of $\alpha$ with $\alpha \geq 2$ satisfy monotonicity and so for these values of $\alpha$ the R\'{e}nyi reflected entropies are correlation measures. Nothing is currently known for non-integer values of $\alpha$ in the range $[2, \infty)$; we did not find any counterexamples to monotonicity in this range. We also wish to emphasize that the papers \cite{dutta2021canonical, bueno2020reflected-2} established monotonicity of reflected entropy for certain states in free and holographic quantum field theories, and that \cite{bueno2020reflected-2, camargo2021long} established a relationship in quantum field theory between the long-distance behavior of reflected entropy and that of mutual information, which is a genuine measure of correlations. Consequently, it is entirely possible that reflected entropy is generally useful as a measure of correlations in physical states of continuum theories.

\section{Counterexamples}

\begin{figure}[h]
	\centering
	\includegraphics[scale=0.5]{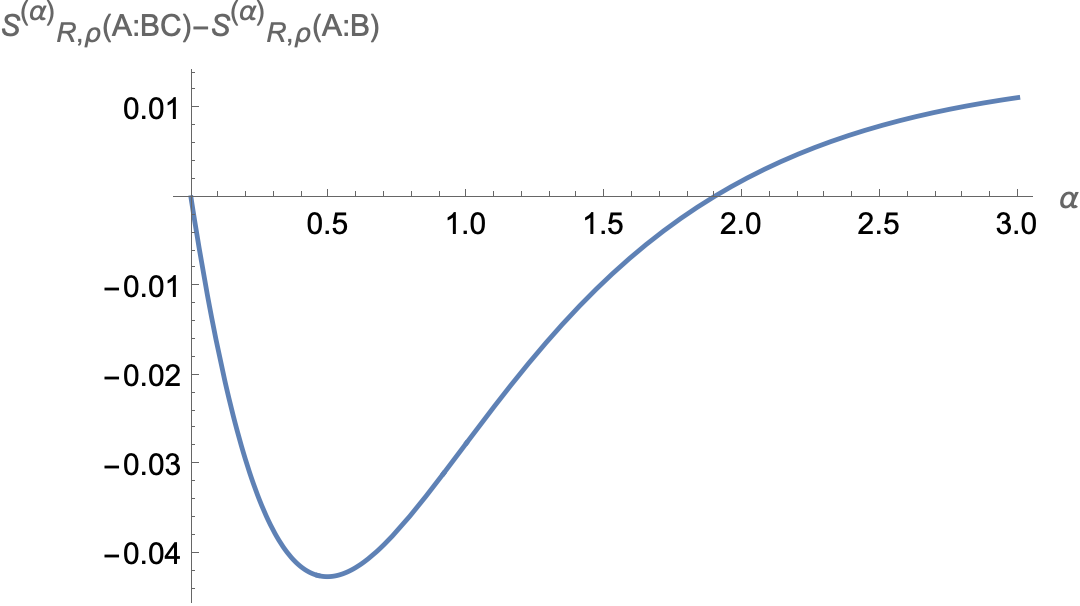}
	\caption{A graph of the difference of the $\alpha$-th R\'{e}nyi reflected entropies $S_{R, \rho}^{(\alpha)}(A:BC)$ and $S_{R,\rho}^{(\alpha)}(A:B)$ for the $\beta=10$ case of the density matrix given in equation \eqref{eq:counterexample}. It is negative for $\alpha=0$ up to around $\alpha=1.9,$ indicating that the R\'{e}nyi reflected entropies are not correlation measures in this range.}
	\label{fig:counterexample}
\end{figure}

We consider the tripartite Hilbert space $\H_{A} \otimes \H_B \otimes \H_C$ where $\H_A$ and $\H_B$ have complex dimension 3, and $\H_C$ has complex dimension 2.
We will denote orthonormal bases of the three spaces by
\begin{gather*}
	\{\ket{0}_{A}, \ket{1}_{A}, \ket{2}_{A}\}, \\
	\{\ket{0}_{B}, \ket{1}_{B}, \ket{2}_{B}\}, \\
	\{\ket{0}_{C}, \ket{1}_{C}\},
\end{gather*}
and will suppress the $A, B, C$ indices.
Our counterexamples are labeled by a single parameter $\beta,$ and are given by the expression
\begin{align} \label{eq:counterexample}
	\rho_{ABC}
		& = \frac{1}{4 \beta + 2} \Bigl( \beta \bigl( \ketbra{000} + \ketbra{110} + \ketbra{200} + \ketbra{210} \bigr) \nonumber \\
		& \quad + \ketbra{020} + \ketbra{121} \Bigr).
\end{align}
It is a straightforward exercise to compute the reflected density operators $\phi^{(AB)}$ and $\phi^{(ABC)}$ on the Hilbert space $\H_A \otimes \H_A^*.$
They are given by
\begin{align}
	\phi^{(ABC)}
		& = \frac{1}{4 \beta + 2} \Bigl( (1 + \beta) \ketbra{00} + \beta \bigl(\ketbra{00}{22} + \ketbra{22}{00} \bigr) + (1 + \beta) \ketbra{11} \nonumber \\
		& \quad + \beta \bigl(\ketbra{11}{22} + \ketbra{22}{11} \bigr) + 2 \beta \ketbra{22} \Bigr)
\end{align}
and
\begin{align}
	\phi^{(AB)}
		& = \phi^{(ABC)} + \frac{1}{4 \beta + 2} \Bigl(\ketbra{00}{11} + \ketbra{11}{00}\Bigr). 
\end{align}
The eigenvalues of these operators can be found analytically.
Each one has six degenerate zero-eigenvalues, and nonzero eigenvalues given by
\begin{align}
	\operatorname{Eigenvalues}(\phi^{(ABC)})
		& = \left\{ \frac{1+\beta}{4 \beta + 2}, \frac{1+3\beta + \sqrt{1 - 2 \beta + 9 \beta^2}}{2(4 \beta + 2)}, \frac{1+3\beta - \sqrt{1 - 2 \beta + 9 \beta^2}}{2(4 \beta + 2)} \right\}, \\
		\operatorname{Eigenvalues}(\phi^{(AB)})
	& = \left\{ \frac{\beta}{4 \beta + 2}, \frac{2+3\beta + \sqrt{4 - 4 \beta + 9 \beta^2}}{2(4 \beta+2)}, \frac{2+3\beta - \sqrt{4 - 4 \beta + 9 \beta^2}}{2(4 \beta + 2)} \right\}.
\end{align}
With these expressions, one can write down formulas for the R\'{e}nyi reflected entropies using equation \eqref{eq:renyi-reflected}, and compute them exactly for any $\alpha$ and any $\beta.$
Figure \ref{fig:counterexample}, for example, shows a plot of $S_{R, \rho}^{(\alpha)}(A:BC) - S_{R, \rho}^{(\alpha)}(A:B)$ for $\beta=10$ in the range $\alpha \in (0, 3).$
The curve is negative from $\alpha=0$ to a value slightly less than $\alpha=1.9,$ indicating that monotonicity of R\'{e}nyi reflected entropies fails in this regime.
This constitutes a counterexample to the conjecture that reflected entropy is a correlation measure, since reflected entropy is given by the $\alpha \to 1$ limit of the R\'{e}nyi family.

In fact, by changing $\beta,$ we can find violations of monotonicity for any R\'{e}nyi reflected entropy with $\alpha \in (0, 2).$
It is easy to check numerically that as one increases $\beta,$ the point where the curve crosses from negative to positive values gets closer and closer to $\alpha=2.$
We will show analytically that for any fixed $\alpha \in (1, 2),$ there exists some value of $\beta$ for which $S_{R,\rho}^{(\alpha)}(A:BC) - S_{R,\rho}^{(\alpha)}(A:B)$ is negative.
Since figure \ref{fig:counterexample} gives an example where this difference is negative for the entire range $0 < \alpha \leq 1,$ the analytic argument for $1 < \alpha < 2$ completes the proof that R\'{e}nyi reflected entropies are not correlation measures in the range $\alpha \in (0, 2).$

We may write the difference as
\begin{equation}
	S_{R, \rho}^{(\alpha)}(A:BC) - S_{R, \rho}^{(\alpha)}(A:B)
		= \frac{1}{1-\alpha} \log \frac{\tr((\phi^{(ABC)})^\alpha)}{\tr((\phi^{(AB)})^\alpha)}.
\end{equation}
In the regime $1 < \alpha < 2,$ the coefficient $1/(1-\alpha)$ is negative, so the overall quantity is negative when the argument of the logarithm is greater than 1, i.e., when we have $\tr((\phi^{(ABC)})^\alpha) > \tr((\phi^{(AB)})^\alpha).$
For fixed $\alpha > 1$ and large $\beta,$ it is easy to verify the asymptotic approximations\footnote{The notation $o(1/\beta)$ means terms strictly smaller than $1/\beta$ in the $\beta \to \infty$ limit; this includes terms like $\beta^{-\alpha}$ and $\beta^{-2}.$}
\begin{equation} \label{eq:asymptotic-ABC}
	\tr((\phi^{(ABC)})^\alpha)
		= \frac{1 + 3^{\alpha}}{4^{\alpha}} - \frac{\alpha}{\beta} \frac{7 \cdot 3^{\alpha}-9}{18 \cdot 4^{\alpha}} + o(1/\beta)
\end{equation}
and
\begin{equation} \label{eq:asymptotic-AB}
	\tr((\phi^{(AB)})^\alpha)
		= \frac{1 + 3^{\alpha}}{4^{\alpha}} - \frac{\alpha}{\beta} \frac{5 \cdot 3^{\alpha}+9}{18 \cdot 4^{\alpha}} + o(1/\beta).
\end{equation}
One then simply checks that for any fixed $\alpha$ in the range $1 < \alpha < 2$, the number $5 \cdot 3^{\alpha} + 9$ exceeds $7 \cdot 3^{\alpha} - 9,$ which implies that for sufficiently large $\beta$ equation \eqref{eq:asymptotic-ABC} will exceed equation \eqref{eq:asymptotic-AB}, implying that the $\alpha$-th R\'{e}nyi reflected entropy is not monotonically decreasing under partial trace.

\acknowledgments{We thank Geoff Penington for helpful discussions, and Dan Eniceicu for introducing JS and ML. PH is supported by ARO (award W911NF2120214), CIFAR, and the Simons Foundation. JS is supported by the Templeton Foundation via the Black Hole Initiative.}

\bibliographystyle{JHEP}
\bibliography{bibliography}

\end{document}